\documentclass[aps,PRB,reprint,superscriptaddress,preprintnumbers]{revtex4-2}
\usepackage[T1]{fontenc}
\usepackage{times}
\usepackage{graphicx,color}
\usepackage{amsfonts,amsmath,amssymb,amsbsy}
\usepackage{ulem}

\begin{document}

\title{Nonadiabatic nonlinear non-Hermitian quantized pumping}
\author{Motohiko Ezawa}
\affiliation{Department of Applied Physics, The University of Tokyo, 7-3-1 Hongo, Tokyo
113-8656, Japan}
\author{Natsuko Ishida}
\affiliation{Research Center for Advanced Science and Technology, The University of
Tokyo, 4-6-1 Komaba, Tokyo 113-8656, Japan}
\author{Yasutomo Ota}
\affiliation{Research Center for Department of Applied Physics and Physico-Informatics,
Keio University, 3-14-1 Hiyoshi, Japan}
\author{Satoshi Iwamoto}
\affiliation{Research Center for Advanced Science and Technology, The University of
Tokyo, 4-6-1 Komaba, Tokyo 113-8656, Japan}

\begin{abstract}
We analyze a quantized pumping in a nonlinear non-Hermitian photonic system
with nonadiabatic driving. The photonic system is made of a waveguide array,
where the distances between adjacent waveguides are modulated. It is
described by the Su-Schrieffer-Heeger model together with a saturated
nonlinear gain term and a linear loss term. A topological interface state
between the topological and trivial phases is stabilized by the combination
of a saturated nonlinear gain term and a linear loss term. We study the
pumping of the topological interface state. We define the 
transfer-speed ratio $\omega /\Omega $ \ by the ratio of the pumping speed $%
\omega $ of the center of mass of the wave packet to the driving speed $%
\Omega $\ of the topological interface. It is quantized as $\omega
/\Omega =1$ in the adiabatic limit. It remains to be quantized for
slow driving even in the nonadiabatic regime, which is a nonadiabatic
quantized pump. On the other hand, there is almost no pump for fast driving.
We find a transition in pumping as a function of the driving speed.
\end{abstract}

\date{\today }
\maketitle

\section{Introduction}

Topological insulator is a prominent idea in contemporary physics\cite%
{Hasan,Qi}. A typical example is a quantum Hall effect or a Chern insulator
in a two-dimensional system described by the Chern number. The Thouless pump
is a dynamical counter part of a Chern insulator\cite%
{Thouless,Niu84,Niu,Citro}, where the Chern number is defined in the
space-time variable. A pumped charge per one cycle is quantized. Especially,
a topological-edge state is pumped in quasicrystal model\cite{Kraus,Verbin},
Rice-Mele model\cite{RWang}, and the Su-Schrieffer-Heeger (SSH) model\cite%
{LonghiPump}.

Photonic systems provide us with an ideal playground of topological physics%
\cite%
{Raghu,KhaniPhoto,Hafe2,Rech,Hafezi,WuHu,KhaniSh,LuRev,OzawaRev,OtaRev,IwamotoRev}%
. Various topological phases are realized in photonic crystal by modulating
the hopping amplitude spatially. A simplest example is the SSH model\cite%
{Jean,Parto,Zhao,Han,Ota18}. Especially, a large area topological interface
laser is theoretically proposed by using the topological interface state of
the SSH model\cite{Ishida,SUSYLaser}. The Thouless pumping is realized by
using spatially modulated waveguides\cite{Kraus,Kraus2,Ke,Verbin,Zilber},
where the hopping amplitude between waveguides are spatially modulated by
modulating the distances between the adjacent waveguides. Dynamics is
governed by the Schr\"{o}dinger equation\cite{Krivo,Longhi}, where the
direction $z$ of the waveguide\ acts as time $t$. Nonlinear Thouless pumping
has been studied in photonic systems\cite{Jurg,Jurg2,Mosta,QFu,QFu2}.
Recently, pumping by modulating the topological interface state of the SSH
model is proposed in a linear Hermitian system\cite{Yuang}, which is not the
Thouless pumping.

Non-Hermicity\cite{LFeng,Weimann,GanaRev} and nonlinearity\cite%
{Ley,Zhou,Malzard,SmiRev,NLPhoto} naturally arises in topological photonics,
which has expanded the field of topological physics starting from condensed
matter physics. Photon loss is effectively well described by a non-Hermitian
loss term. On the other hand, the gain has a saturation, which is described
by a nonlinear term. Stable laser emission occurs in the presence of both of
these terms\cite{Harrari,Bandres}. The interplay of non-Hermicity and
nonlinearity is interesting\cite{Hass,MalzardOpt,EzawaLaser}.

The Thouless pumping is valid only in the adiabatic limit. The pumped charge
is not quantized in the nonadiabatic regime\cite{Privi,Tuloup}. It is
interesting if a pumping can be quantized in the nonadiabatic regime in
general.

In this paper, we study a nonadiabatic quantized pumping in a nonlinear
non-Hermitian photonic system. The basic structure is described by the SSH
model, which has the topological and trivial sectors. The interface state
emerges at the interface between these two sectors. We investigate the
pumping of the state by changing the pumping velocity, and find a transition
to occur between a quantized charge pumping and no pumping. We define the 
transfer-speed ratio by the ratio $\omega /\Omega $, where $\omega $\
is the pumping speed of the wave packet while $\Omega $\ is the driving
speed of the topological interface. The transfer-speed ratio 1 means
that the wave packet perfectly follows the driving. In slow pumping, the
wave function of the interface state perfectly follows the modulation, and
hence the transfer-speed ratio is 1. On the other hand, in fast
pumping the wave function collapses because it cannot follow the modulation,
and hence, the transfer-speed ratio is zero.

\begin{figure}[t]
\centerline{\includegraphics[width=0.48\textwidth]{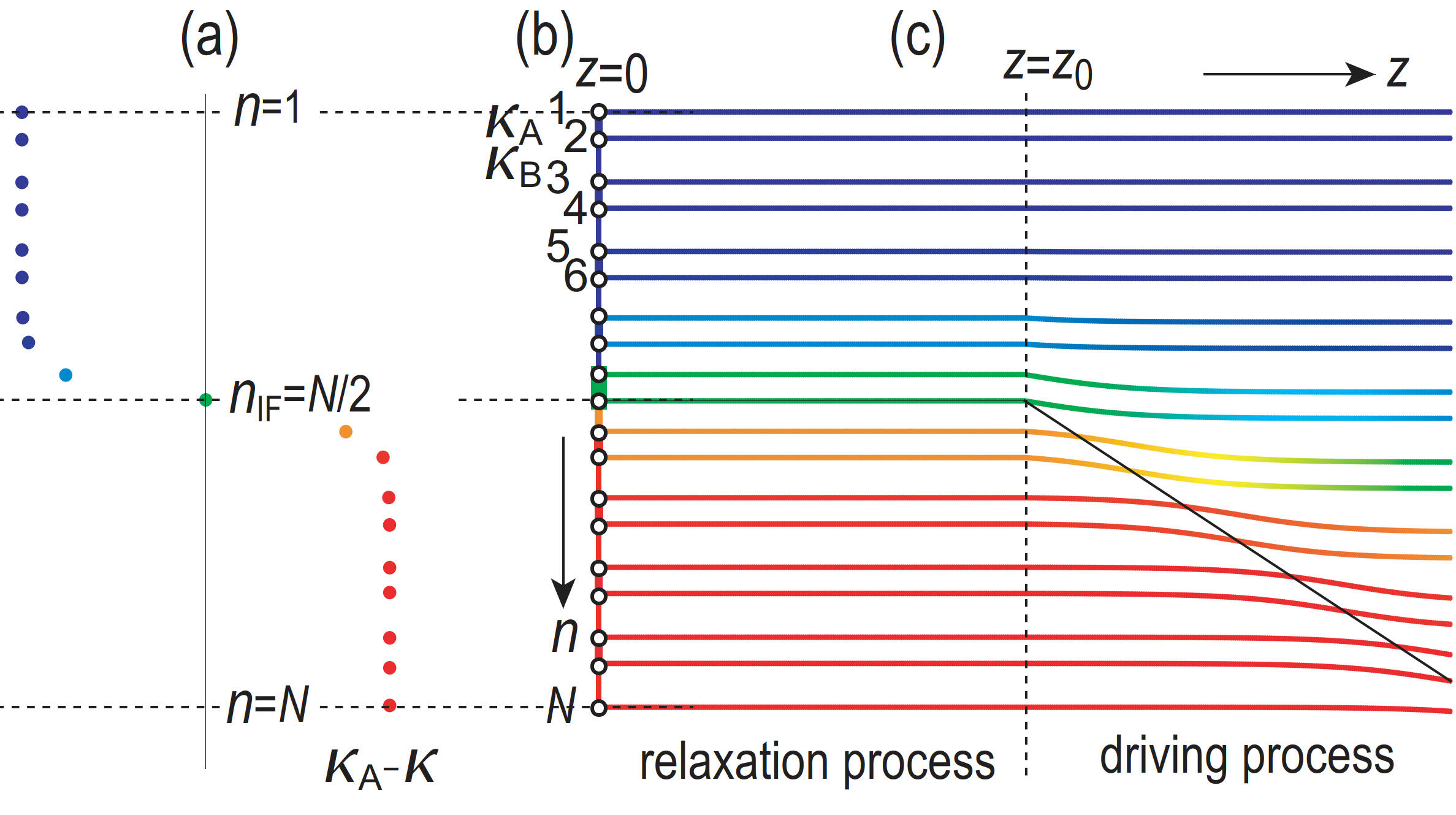}}
\caption{Illustration of modulated photonic waveguide. (a) The modulation is
given to the hopping parameter $\protect\kappa _{A}$ according to Eq.(%
\protect\ref{EqA}) with $n_{\text{IF}}(z)=N/2$ at $z=0$. (b) The basic
structure is the SSH chain made of $N$ sites. (c) The modulation is given in
the $n$-$z$ plane by Eq.(\protect\ref{EqA}) with $n_{\text{IF}%
}(z)=N/2+\Omega (z-z_{0})$, as indicated by a black line. Here, $N=20$.}
\label{FigIllust}
\end{figure}

\begin{figure*}[t]
\centerline{\includegraphics[width=0.95\textwidth]{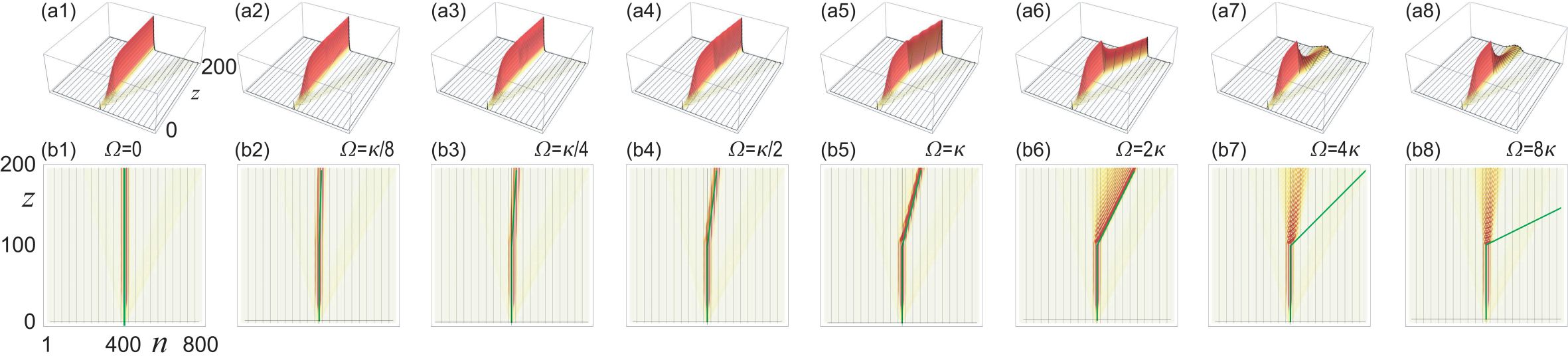}}
\caption{Wave packet in quench dynamics in bird's eye view and in the ($n,z$%
) plane. (a1), (b1) $\Omega =0$; (a2), (b2) $\Omega =\protect\kappa /8$;
(a3), (b3) $\Omega =\protect\kappa /4$; (a4), (b4) $\Omega =\protect\kappa %
/2 $; (a5), (b5) $\Omega =\protect\kappa $; (a6), (b6) $\Omega =2\protect%
\kappa $; (a7), (b7) $\Omega =4\protect\kappa $; (a8), (b8) $\Omega =8%
\protect\kappa $. We have used a chain with a length $N=800$. We have set $%
\protect\eta =10,\protect\xi =20$, $\protect\lambda =0.5,\protect\gamma =0.1$
and $\protect\chi =1$. The evolution is calculated in the range $0\leq z\leq
200$, where $z_{0}=100$. Green lines represent the interface line (\protect
\ref{Modul}).}
\label{FigDynamics}
\end{figure*}

\section{Model}

We consider an array of spatially modulated waveguides\cite%
{Kraus,Kraus2,Verbin,Zilber,Rech,Gara}, as illustrated in Fig.\ref{FigIllust}%
. The basic structure is described by the SSH chain made of $N$ sites as in
Fig.\ref{FigIllust}(b), which is a bipartite system. We take $N$ to be an
even integer. The hopping parameter $\kappa _{A,n}$\ has a site dependence,
while $\kappa _{B}$ does not. They are given by%
\begin{equation}
\kappa _{A,n}\left( z\right) =\kappa \left( 1+\lambda \tanh \frac{n-n_{\text{%
IF}}\left( z\right) }{\xi }\right) ,\quad \kappa _{B}=\kappa ,  \label{EqA}
\end{equation}%
where $n_{\text{IF}}\left( z\right) $ is the position of the interface as a
function of $z$; $\lambda >0$ and $\xi >0$\ represent the interface
modulation amplitude and the interface width, respectively. Small (large) $%
\xi $\ represents a sharp (smooth) interface. The function $\kappa
_{A,n}-\kappa $\ is illustrated in Fig.\ref{FigIllust}(a) with the choice of 
$n_{\text{IF}}\left( z\right) =N/2$. We set $\xi =20$, $\lambda =0.5$, $%
\kappa =1$, $\gamma =0.1$, $\chi =1$\ and $\eta =10$\ in numerical studies,
unless otherwise stated.

The system is governed by\cite{Harrari} 
\begin{equation}
i\frac{d\psi _{n}}{dz}=\sum_{nm}M_{nm}\psi _{m}-i\gamma \left( 1-\chi \frac{%
\left( 1-\left( -1\right) ^{n}\right) /2}{1+\left\vert \psi _{n}\right\vert
^{2}/\eta }\right) \psi _{n},  \label{Master}
\end{equation}%
with a site-dependent hopping matrix representing the SSH model,%
\begin{align}
& M_{nm}\left( z\right)  \notag \\
& =\left( 
\begin{array}{cccccc}
0 & \kappa _{A,n}\left( z\right) & 0 & 0 & 0 & \cdots \\ 
\kappa _{A,n}\left( z\right) & 0 & \kappa _{B} & 0 & 0 & \cdots \\ 
0 & \kappa _{B} & 0 & \kappa _{A,n}\left( z\right) & 0 & \cdots \\ 
0 & 0 & \kappa _{A,n}\left( z\right) & 0 & \kappa _{B} & \ddots \\ 
0 & 0 & 0 & \kappa _{B} & 0 & \ddots \\ 
\vdots & \vdots & \vdots & \ddots & \ddots & \ddots%
\end{array}%
\right) ,  \label{HoppiMatrix}
\end{align}%
where $\psi _{n}$ is the amplitudes at the site $n$, where $n=1,2,3,\cdots
,N $; $\gamma $ represents the constant loss in each waveguide; $\gamma \chi 
$ represents the amplitude of the optical gain via stimulated emission
induced only at the odd site; $\eta $ represents the saturation parameter of
nonlinear gain\cite{Harrari}. All these parameters are positive. The system
turns out to be a linear model in the limit $\eta \rightarrow \infty .$ On
the other hand, $\gamma $\ controls the non-Hermicity, where the system is
Hermitian for $\gamma =0$. In Eq.(\ref{Master}) we measure $z$ in units of $%
1/\kappa $ and the loss parameter $\gamma $ in units of $\kappa $, where $%
\kappa $ is defined in Eq.(\ref{EqA}).

The explicit equations for a finite chain with length $N$\ follow from Eq.(%
\ref{Master}) as%
\begin{align}
i\frac{d\psi _{2n-1}}{dz}=& \kappa _{B}\psi _{2n-2}+\kappa _{A,n}\left(
z\right) \psi _{2n}  \notag \\
& -i\gamma \left( 1-\frac{\chi }{1+\left\vert \psi _{2n-1}\right\vert
^{2}/\eta }\right) \psi _{2n-1},  \label{SSH1} \\
i\frac{d\psi _{2n}}{dz}=& \kappa _{B}\psi _{2n+1}+\kappa _{A,n}\left(
z\right) \psi _{2n-1}-i\gamma \psi _{2n}.  \label{SSH2}
\end{align}%
We solve this set of equations together with the boundary condition%
\begin{equation}
\psi _{n}\left( z=0\right) =\delta _{n,n_{\text{IF}}}.  \label{IniCon}
\end{equation}

It is convenient to regard $z$ as time $t$. Then, Eqs.(\ref{SSH1}) and (\ref%
{SSH2}) are the Schr\"{o}dinger equations describing a quench dynamics
starting from the interface site by giving an input to it with the initial
condition (\ref{IniCon}). We consider a case where%
\begin{equation}
n_{\text{IF}}\left( z\right) =N/2  \label{ModulC}
\end{equation}%
for the relaxation process $z<z_{0}$, and 
\begin{equation}
n_{\text{IF}}\left( z\right) =N/2+\Omega \left( z-z_{0}\right)  \label{Modul}
\end{equation}%
for the driving process $z>z_{0}$. See Fig.\ref{FigIllust}(c) for the
modulation of the hopping parameters with this choice of Eqs.(\ref{ModulC})
and (\ref{Modul}). In what follows, we may occasionally regard $z$ as time $%
t $. The pumping is said adiabatic only for $\Omega \ll \kappa $, which
means that any finite driving is nonadiabatic.

\section{Nonadiabatic regime}

We start with a review of the stationary solution in the case of $\Omega =0$%
, where the interface $n_{\text{IF}}\left( z\right) $ is given by the
constant (\ref{ModulC}) in the hopping parameter (\ref{EqA}) for all $z$.
This case was studied previously, where the stationary solution is found to
be\cite{SUSYLaser}%
\begin{align}
\Psi _{A}\left( n\right) & \varpropto \exp \left[ -\frac{\kappa \lambda }{%
2\xi }\left( 1+c_{2}\right) (n-n_{\text{IF}}\left( z\right) )^{2}\right] , 
\notag \\
\Psi _{B}\left( n\right) & \varpropto -i\frac{x}{\eta }\frac{c_{2}\kappa
\lambda }{\xi }\exp \left[ -\frac{\kappa \lambda }{2\xi }\left(
1+c_{2}\right) (n-n_{\text{IF}}\left( z\right) )^{2}\right] ,  \label{JR}
\end{align}%
together with (\ref{ModulC}) for all $z$, where $c_{2}$ is a certain
constant: See Eq.(78) in the reference\cite{SUSYLaser}. This is a
generalization of the Jackiw-Rebbi solution to a nonlinear non-Hermitian
model. The quench dynamics was also examined, where the wave packet spreads
from the delta function (\ref{IniCon}) and reaches the stational
distribution (\ref{JR}) as in Fig.\ref{FigDynamics}(a1) and (b1). Namely,
the interface state (\ref{JR}) is formed in the relaxation process ($z<z_{0}$%
).

We investigate the driving process, where the interface $n_{\text{IF}}\left(
z\right) $ is modulated as in Eq.(\ref{Modul}) for $z>z_{0}$. In the
adiabatic approximation ($\Omega \simeq 0$), the wave packet is expected to
perfectly follow the interface state (\ref{JR}) together with (\ref{Modul}).
This is indeed the case, as numerically checked in Fig.\ref{FigDynamics}(a2)
and (b2) for $\Omega =\kappa /8$.

What is unexpected is the numerical result for larger values of $\Omega $.
For slow driving $\Omega <\kappa $, the state maintains its form as in Fig.%
\ref{FigDynamics}(a1)$\sim $(a5), whose center perfectly follows the sample
modulation (\ref{Modul}),\ as shown by the green line in Fig.\ref%
{FigDynamics}(b1)$\sim $(b5). On the other hand, for fast driving $\Omega
>\kappa $, the wave packet collapses and spreads, as shown in Fig.\ref%
{FigDynamics}(b6)$\sim $(a8) and (b6)$\sim $(b8).

\begin{figure}[t]
\centerline{\includegraphics[width=0.48\textwidth]{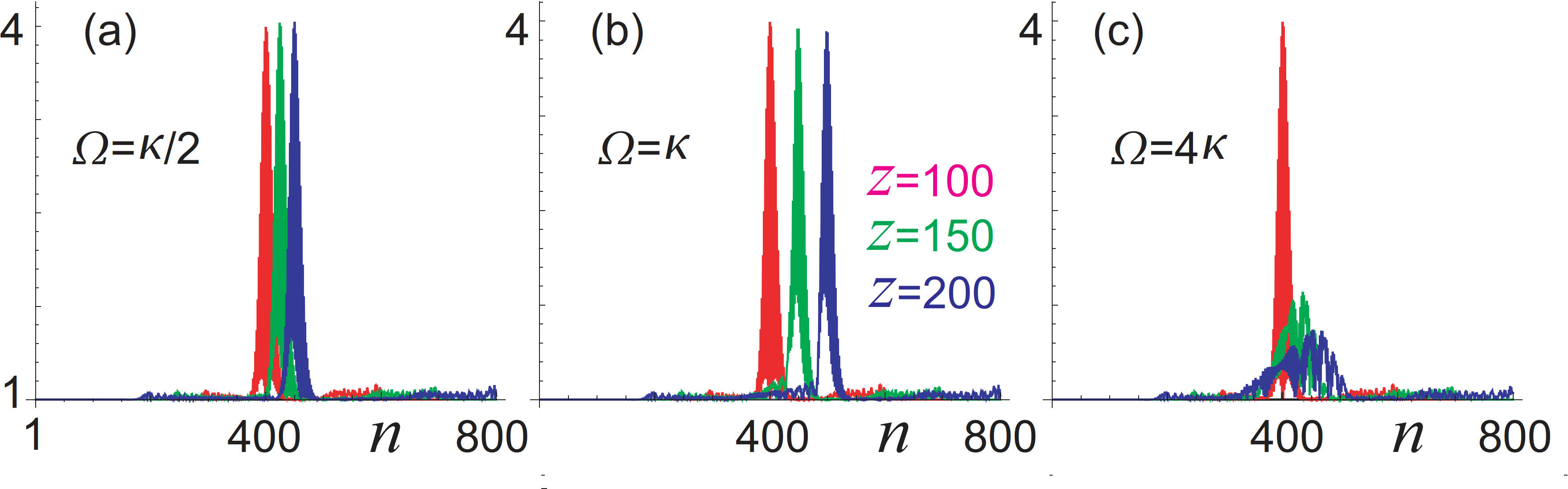}}
\caption{Spatial profile of the wave packet for (a) slow driving $\Omega =%
\protect\kappa /2$, (b) the critical driving $\Omega =\protect\kappa $ and
(c) fast driving $\Omega =4\protect\kappa $. The red curves are the spatial
profile at $z=100$, the green curves are that at $z=150$ and the blue curves
are that at $z=200$, where $z_{0}=100$.}
\label{FigCut}
\end{figure}

We show the snapshot at $z=100$, $150$ and $200$ for $\Omega =\kappa /2$, $1$
and $4$ in the system with $z_{0}=100$ in Fig.\ref{FigCut}. The wave packet
does not change its form for $\Omega =\kappa /2$ and $1$,\ as shown in Fig.%
\ref{FigCut}(a) and (b), which leads to the quantized pumping. On the other
hand, the wave packet is collapsed for $\Omega =4\kappa $ as shown in Fig.%
\ref{FigCut}(c), where the pumping does not occur.

\begin{figure}[t]
\centerline{\includegraphics[width=0.48\textwidth]{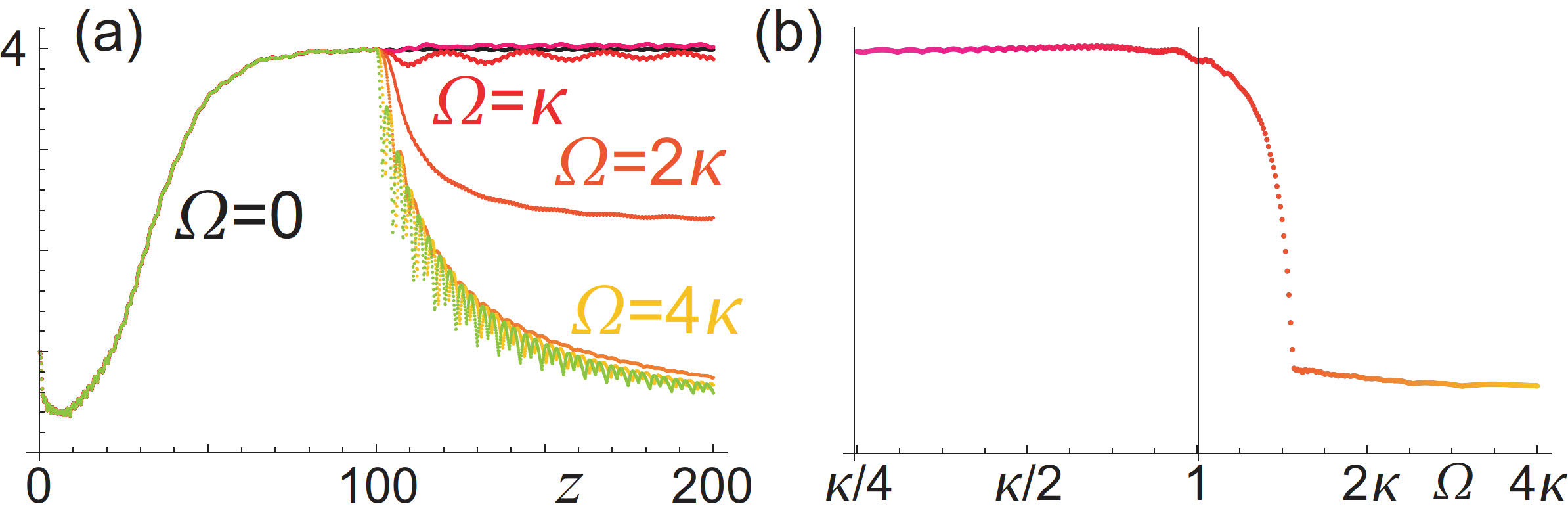}}
\caption{ (a) Maximum value of the wave packet as a function of $z$. (b) The
maximum value of the wave packet at $z=200$ as a function of $\Omega /%
\protect\kappa $ in the logarithmic scale (log$_{2}\Omega /\protect\kappa $).}
\label{FigMax}
\end{figure}

\begin{figure}[t]
\centerline{\includegraphics[width=0.48\textwidth]{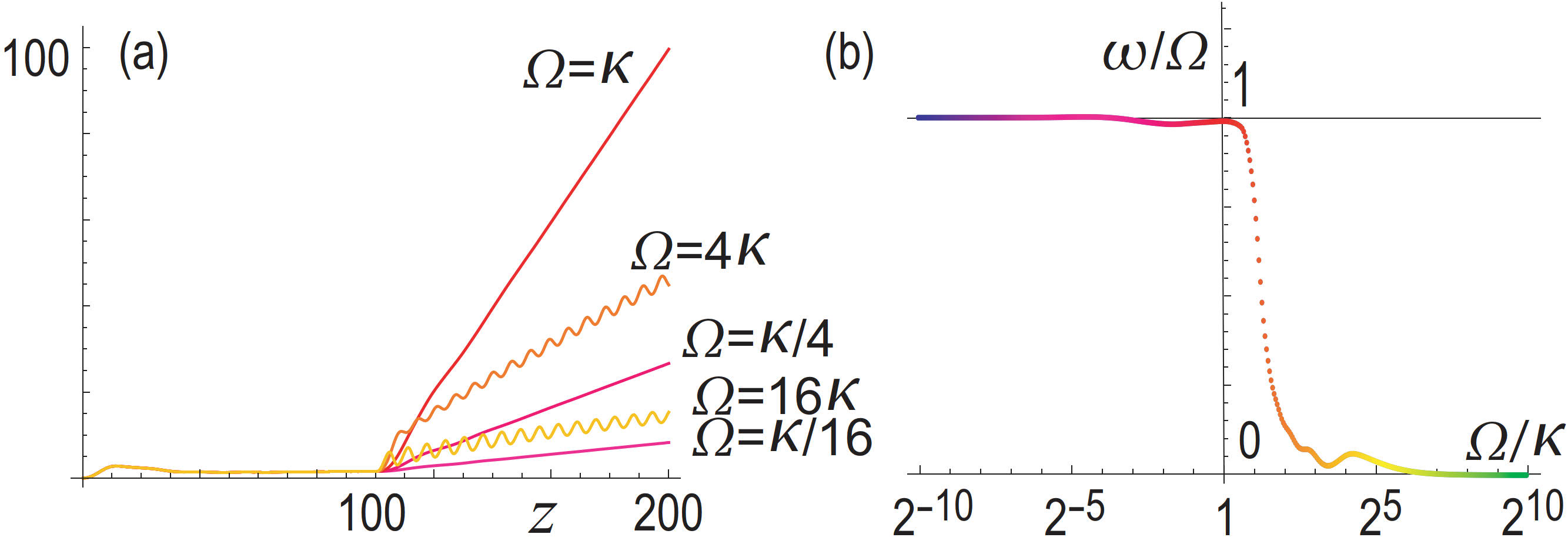}}
\caption{(a) Evolution of the mean position $\left\langle x\left( z\right)
\right\rangle $. The pumping velocity $\protect\omega $ is determined from
the slop of the line for each $\Omega /\protect\kappa $. The horizontal axis
is $z$ with $0\leq z\leq 200$. (b) Ratio $\protect\omega /\Omega $ as a
function of $\Omega $. The inset is an enlarged curve in the vicinity of $%
(\Omega /\protect\kappa ,\protect\omega /\Omega )=(1,1)$. The horizontal
axis is in the logarithmic scale (log$_{2}\Omega /\protect\kappa $). }
\label{FigCritical}
\end{figure}

We show the $z$\ evolution of the maximum value of the wave packet in Fig.%
\ref{FigMax}(a). The maximum value increases in the relaxation process and
reaches a stational value. It maintains as it is for slow driving ($\Omega
<\kappa $). On the other hand, it decreases as soon as fast driving ($\Omega
>\kappa $) is started at $z=100$. We plot the maximum value at $z=200$\ in
Fig.\ref{FigMax}(b). It is constant for slow driving and decreases for fast
driving.

We have shown the existence of the critical value $\Omega =\kappa $ for the
driving velocity. In order to determine this critical value more clearly, we
calculate the expectation value of the mean position by%
\begin{equation}
\left\langle x\left( z\right) \right\rangle \equiv \frac{\sum_{n}\left( n-n_{%
\text{IF}}\left( 0\right) \right) \left\vert \psi _{n}\left( z\right)
\right\vert ^{2}}{\sum_{n}\left\vert \psi _{n}\left( z\right) \right\vert
^{2}}.
\end{equation}%
We show the $z$\ evolution of $\left\langle x\left( z\right) \right\rangle $%
\ for various $\Omega $\ in Fig.\ref{FigCritical}(a). It is almost zero in
the relaxation process ($z<z_{0}$). On the other hand, it increases almost
linearly in the driving process ($z>z_{0}$). Namely, $\left\langle x\left(
z\right) \right\rangle $ is well approximated by%
\begin{equation}
\left\langle x\left( z\right) \right\rangle =\omega \left( z-z_{0}\right) .
\end{equation}%
We define the pumping velocity $\omega $ at each $\Omega $\ by this formula.

We then calculate numerically the ratio $\omega /\Omega $ of the pumping
speed $\omega $\ to the driving speed $\Omega $. We plot $\omega /\Omega $
as a function of $\Omega $ in Fig.\ref{FigCritical}(b), where it is found
that the pumping speed is almost identical to the driving speed, i.e., $%
\omega =\Omega $ for $\Omega \leq \kappa $. It means that the wave packet
follows the modulation $n_{\text{IF}}\left( z\right) $. On the other hand, $%
\omega $ decreases as the increase of $\Omega $. It means that the
modulation is too fast for the wave packet to follow it. The transition
occurs near $\Omega \simeq \kappa $, where the system is nonadiabatic.

\begin{figure}[t]
\centerline{\includegraphics[width=0.48\textwidth]{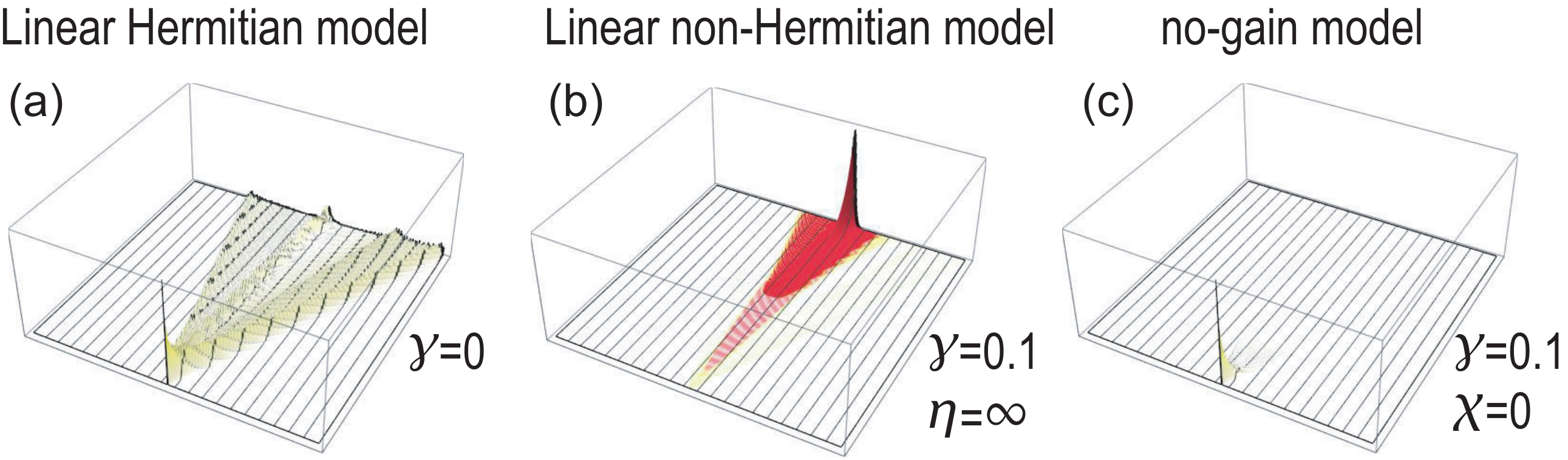}}
\caption{(a) Linear Hermitian model ($\protect\gamma =0$), (b) Linear
non-Hermitian model ($\protect\gamma =0.1$, $\protect\eta =\infty )$. and
(c) non-Hermitian with the loss term and without the gain term ($\protect%
\gamma =0.1$, $\protect\chi =0)$. We have set $\Omega =\protect\kappa /2$.}
\label{FigLinear}
\end{figure}

The quantized pumping in the nonadiabatic regime requires both non-Hermicity
and nonlinearity. The wave packet rapidly collapses for the Hermitian system
($\gamma =0$) as in Fig.\ref{FigLinear}(a). This is because the delta
function contains components not only from the topological interface state
but also from many bulk states, where the bulk states spreads rapidly over
the sample\cite{NLPhoto,SUSYLaser}. The wave packet exponentially grows and
does not reach the stationary solution for the linear non-Hermitian system (%
$\gamma \neq 0,$ $\eta =\infty )$ as in Fig.\ref{FigLinear}(b). The wave
packet rapidly decreases if there is the loss term without the gain term ($%
\gamma \neq 0,\chi =0)$ as in Fig.\ref{FigLinear}(c).

\section{Discussion}

We have shown that the nonlinear non-Hermitian pump is quantized even in the
nonadiabatic regime. The combination of the saturated gain and linear loss
greatly stabilizes the wave packet, and hence, it is robust even for
nonadiabatic driving. It is highly contrasted to the Thouless pump, which is
broken in the nonadiabatic regime\cite{Privi,Tuloup}. Our work provides an
example that the interplay among non-Hermicity and nonlinearity gives an
intriguing phenomena in the nonadiabatic regime.

We compare the maximum driving speed $\Omega_{\text{max}}$ in the
nonadiabatic regime with those in the previous work\cite{LonghiPump}, where
the system is Hermitian and linear. It is of the order of
$\Omega_{\text{max}}/\kappa=30/5000\eqsim 0.006$ in the Thouless pumping
based on the Rice-Mele model, and it is of the order of
$\Omega_{\text{max}}/\kappa=31/400\eqsim 0.075$ in the pumping of the
topological interface state. On the other hand, it is of the order of
$\Omega_{\text{max}}/\kappa=1$ in the present mechanism. Hence, the ratio is
much faster than the previous ones. This is because that the wave packet is
stabilized by the combination of the non-Hermicity and nonlinearity.

Femto-second laser writing waveguide\cite{Rech,Davis,SzaFemto} or
semiconductor waveguide\cite{Foresi,Foster,Sun} are used in experiments. 
Typical distance between waveguides is 15$\mu m$ (5$\mu m$) and the length of the waveguide is 500mm 
(50$\mu m$) for a
femto-second laser writing\cite{Rech} (semiconductor waveguide\cite{Sun})
waveguide. We have set $N=800$, which corresponds to 12mm for a
femto-second laser writing waveguide and 4mm for a semiconductor waveguide,
in numerical simulations.

M.E is supported by CREST, JST (Grants No. JPMJCR20T2) and Grants-in-Aid for
Scientific Research from MEXT KAKENHI (Grant No. 23H00171). N. Ishida is
supported by the Grants-in-Aid for Scientific Research from MEXT KAKENHI
(Grants. No. JP21J40088).\ Y. Ota is supported by the Grants-in-Aid for
Scientific Research from MEXT KAKENHI (Grants. Nos. 22H01994 and 22H00298).
S. Iwamoto is supported by CREST, JST (Grants No. JPMJCR19T1) and the
Grants-in-Aid for Scientific Research from MEXT KAKENHI (Grants. Nos.
22H00298 and 22H01994).

\end{document}